\documentclass[useAMS,usenatbib]{mn2e}

\usepackage{amssymb,amsmath,upgreek}

\usepackage{epsfig,graphicx,times,subfigure}

\newcommand{\msun}{M$_{\odot}$}

\newcommand{\lsun}{L$_{\odot}$}

\newcommand{\aj}{AJ}

\newcommand{\pasa}{PASA}
\newcommand{\apj}{ApJ}
\newcommand{\mnras}{MNRAS}

\defcitealias{Fra12}{F12}
\defcitealias{Van12}{vD12}

\title[SN 2012aw]
{The disappearance of the progenitor of SN 2012aw in late-time imaging}

\author[M. Fraser]
{Morgan Fraser$^{1}$\thanks{E-mail:mf@ast.cam.ac.uk}
\\ 
$^{1}$Institute of Astronomy, University of Cambridge, Madingley Road, Cambridge, CB3 0HA, UK\\
}

\begin{document}

\date{Submitted to Monthly Notices of the Royal Astronomical Society}

\pagerange{\pageref{firstpage}--\pageref{lastpage}} \pubyear{}

\maketitle

\label{firstpage}

\begin{abstract}
We present new late-time near-infrared imaging of the site of the nearby core-collapse supernova SN 2012aw, confirming the disappearance of the point source identified by \cite{Fra12} and \cite{Van12} as a candidate progenitor in both {\it J} and {\it Ks} filters. We re-measure the progenitor photometry, and find that both the {\it J} and {\it Ks} magnitudes of the source are consistent with those quoted in the literature. We also recover a marginal detection of the progenitor in {\it H}-band, for which we measure {\it H}=$19.67\pm0.40$~mag. Comparing the luminosity of the progenitor to stellar evolutionary models, SN 2012aw appears to have resulted from the explosion of a 12.5$\pm$1.5 \msun\ red supergiant.
\end{abstract}

\begin{keywords}
supernovae: general -- supernovae: individual: SN 2012aw -- stars: massive
\end{keywords}

\section{Introduction}
\label{s1}

Core collapse supernovae (SNe) are luminous and energetic transients associated with the demise of an evolved massive star. They play a critical role in the production and distribution of metals within galaxies \citep[e.g.][]{Nom06}, and regulating star formation and galaxy evolution through feedback \citep[e.g.][]{Sca08}.
Nearby supernovae with identified progenitors can also be used to test the predictions of stellar evolutionary models and understand the connection between the final stages of massive stellar evolution, and the resulting explosion and compact remnant \citep{Heg03,Eld08}.

In the last decade the detection of  progenitor candidates for nearby core-collapse supernovae in archival imaging has become relatively routine \citep[see][for a recent review]{Sma15}. To confirm these candidates requires deep images to be taken after the SN has faded, in which the progenitor candidate should be no longer visible. This technique has been used to verify the progenitors of some nearby SNe \citep{Van13,Mau14}, but has also revealed several erroneous identifications \citep{Mau15}.

SN 2012aw is a nearby, well studied Type II-Plateau (IIP) SN in M95, discovered by \cite{Fag12}. Followup observations of SN 2012aw were presented by \cite{Bos13} and \cite{Dal14}; the SN appears relatively bright and quite similar to SN 1999em, with an ejected $^{56}$Ni mass of 0.06~\msun. 
Both \cite{Fra12} and \cite{Van12} (henceforth \citetalias{Fra12} and  \citetalias{Van12}) identified a progenitor candidate for SN 2012aw in archival {\it Hubble Space Telescope} and ground based near-infrared (NIR) data. The candidate had an apparent magnitude in {\it F814W} ($\sim${\it I}-band) of 23.30$\pm$0.02 mag, however its colour was redder than one might expect for even the coolest red supergiants, with {\it F555W}$-${\it F814W}  between 3.1 and 3.3 mag, leading both \citetalias{Fra12} and \citetalias{Van12} to suggest that the progenitor suffered from significant extinction. \citetalias{Van12} and \citetalias{Fra12} estimated that the progenitor had a zero-age main sequence (ZAMS) mass of 17--18 \msun\ or 14--26 \msun\ respectively. However, \cite{Koc12} pointed out that both of these mass ranges were likely overestimates due to an incorrect treatment of the effects of circumstellar dust. \citeauthor{Koc12} argued that a purely silicate dust composition would be more appropriate for a red supergiant, and that dust emission in the NIR must be considered. Moreover, they point out that scattered photons from a shell of dust around a star will also contribute to its observed flux; resulting in a tendency to over-estimate the effects of extinction on luminosity when using a \cite{Car89} extinction law.

In this Letter, we present new NIR observations obtained three years after the explosion of SN 2012aw, which we use to verify the identification of the progenitor candidate found by both \citetalias{Fra12} and \citetalias{Van12}. We adopt a distance of 9.8$\pm0.2$~Mpc to M95 (based on the average of the \citealp{Fre01} and \citealp{Riz07} distances, derived from Cepheids and the Tip of the Red Giant Branch respectively; these are consistent with the 10 Mpc distance used by \citetalias{Fra12}). We take the foreground (Milky Way) extinction to be $A_V$=0.08 mag; consistent with values used in \citetalias{Fra12}, while in in Sect. \ref{s4} we adopt the values for circumstellar extinction around the progenitor from \cite{Koc12}.

\section{Observations and data reduction}
\label{s2}

New observations of the site of SN 2012aw were obtained with the 8.2~m Very Large Telescope (VLT) + High Acuity Wide field K-band Imager \citep[HAWK-I;][]{Kis08} on the nights of 2015 March 2 and 7. HAWK-I consists of four 2048$\times$2048 pixel detectors, with a combined field of view of 7.5\arcmin$\times$7.5\arcmin, and a pixel scale of 0.106\arcsec/pix. On March 2, M95 was observed using the {\it Ks} filter on MJD 57084.10, at an airmass between 2.0 and 1.6. On March 7, observations were taken on 57089.2 using the {\it J} filter at an airmass of between 1.2 and 1.3.

For both {\it J} and {\it Ks}, four consecutive frames were taken at a single pointing, with the site of SN 2012aw approximately centred on the Q1 chip of HAWK-I, before the telescope was dithered $\sim$5\arcsec\ to a new position. The site of SN 2012aw is $\sim$2\arcmin\ from the centre of M95, and the background at this location is sufficiently sparse that separate off-target sky frames are not required. Each frame consisted of 3$\times$10s integrations (DIT=10~s, NDIT=3 in ESO terminology), which were read out in non-destructive readout mode. A total of 60 frames were obtained in {\it J} and 68 frames in {\it Ks}, to give combined on-source exposure times of 1800~s and 2040~s in {\it J} and {\it Ks} respectively.

All data were reduced using the HAWK-I pipeline, running under the {\sc esorex} environment. A bad pixel mask was constructed by identifying outlier pixels in dark frames. Twilight flat fields for {\it J} and {\it Ks} were created, and were used along with the bad pixel masks to reduce the science frames. The most critical step in the reduction of NIR images is the removal of the sky background; for our data we employed a two-step procedure. A first pass sky image for each frame was created by median combining the 8 images taken at the two preceding dither positions, and the 8 images taken at the two subsequent dither positions. The sky image associated with each frame was then subtracted to give the first pass reduced images, which were then shifted and co-added. The combined first pass image was then used to identify point and extended sources, and create a mask for these. This mask was then shifted as necessary and applied to each frame. The masked images were then used to create an improved sky frame for each image (again using 16 images taken before and after each frame at different dither positions). The refined sky frame was subtracted from each of the original flat-fielded, bad pixel-masked images to give the second pass reduced images. Finally, the second pass reduced images were corrected for geometric distortion using tabulated distortion coefficients, before being combined to give the final reduced image. 

The measured seeing of both the reduced {\it J} and {\it Ks} images was 4.6 pixels (0.5\arcsec). According to the ESO Ambient Conditions Database, both nights were  photometric at the time the observations were taken.

The pre-explosion images used in \citetalias{Fra12} were re-reduced for this work. We considered both {\it Js}-band images from VLT + ISAAC taken on 2000 March 25, and the New Technology Telescope + SOFI {\it H} and {\it Ks} images taken on 2002 March 24. We note that the {\it H}-band images were not analysed in \citetalias{Fra12} as the exposure times are relatively short, however, a careful reduction reveals a source at the location of SN 2012aw. The ISAAC data have a pixel scale of 0.15 pix~arcsec$^{-1}$, while SOFI has a scale of 0.29 pix~arcsec$^{-1}$.

For the SOFI data, a bad pixel mask appropriate to the epoch of the observations was obtained from the SOFI webpages. The reductions were carried out with {\sc iraf}, with the exception of determining the illumination correction frame, for which an {\sc esorex} script was used. Flat field images for both {\it H} and {\it Ks} were derived using a set of dome flats taken both with and without lamp illumination. As the dome flat does not exactly match the illumination of the sky, an illumination correction frame was derived from observations of a bright star which was imaged at 16 different positions on the detector. The bad pixel mask, flat field and illumination correction were then applied to all science frames. A sky frame for each filter was created from the mean of the science frames, using sigma clipping to remove sources. As the images are relatively shallow, it was not found to be necessary to use a two pass sky subtraction. Finally, the images taken in each filter were shifted to a common pixel grid and co-added. 10 {\it K}-band images were used (each of which comprised 10$\times$6~s DIT) for a total exposure time of 600~s, while in {\it H}-band 4 frames were taken with 3$\times$15~s DIT) for 180~s on source. 

The ISAAC {\it Js}-band data were reduced within {\sc esorex}. Flat fields were derived from twilight sky observations, while a bad pixel mask was constructed from dark frames; these were then applied to all science frames. Separate off-target sky frames were available for these observations, and so we used the off-source frames taken immediately prior and subsequent to each science frame to create a sky frame, which was then subtracted from the on-source images. The reduced, sky-subtracted images were then aligned within {\sc iraf} and co-added to produce a single, deep image. The location of SN 2012aw lies very close to the edge of the detector in some of the frames, and so we did not include these in the final stacked image. The combined image is comprised of 5 exposures, each of which has 4$\times$30~s DIT, to give a total time exposure of 600~s.

\section{Analysis}
\label{s3}

\begin{figure}
\includegraphics[width=0.7\linewidth,angle=270]{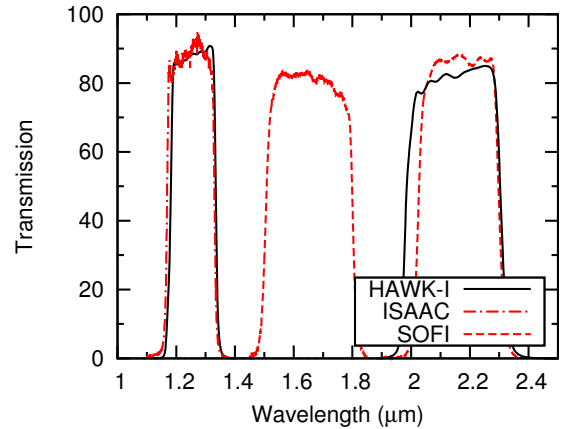}
\caption{Filter transmission curves for ISAAC {\it Js} and HAWK-I {\it J}; SOFI {\it H}; and SOFI {\it Ks} and HAWK-I {\it Ks}}
\label{fig:filter}
\end{figure}

Ideally, the late-time observations of the site of SN 2012aw would be taken with the same instrument and filter as the pre-explosion data. As this was not possible, we tested the effects of mis-matched filters using synthetic photometry. We obtained the throughput curves for the filters used from the ESO webpages, these are shown in Fig. \ref{fig:filter}. We used the throughput curves together with \cite{Cas04} models appropriate to a massive red supergiant and the {\sc pysynphot} package to calculate the expected {\it Js$_\mathrm{ISAAC}$}$-$ {\it J$_\mathrm{HAWK-I}$} and  {\it Ks$_\mathrm{SOFI}$}$-${\it Ks$_\mathrm{HAWK-I}$} zeropoint differences (effectively a colour) for the progenitor. For {\it J}, the largest difference is 0.09~mag, while for {\it Ks} the difference is $<-0.01$~mag.

\begin{figure*}
\centering
\subfigure[ISAAC {\it Js} pre-explosion]{
\includegraphics[width=0.3\textwidth]{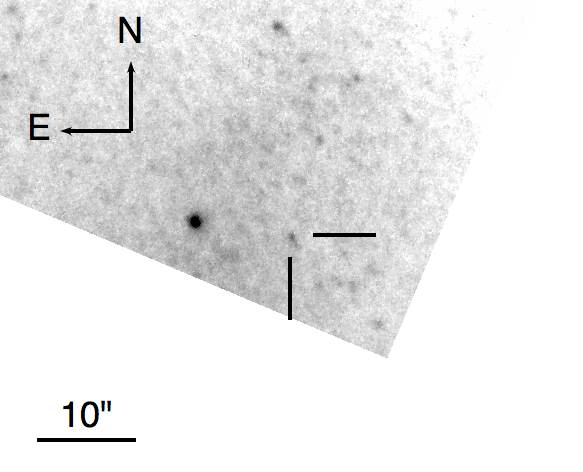}\llap{\raisebox{2.3cm}{\includegraphics[width=2cm,height=2cm]{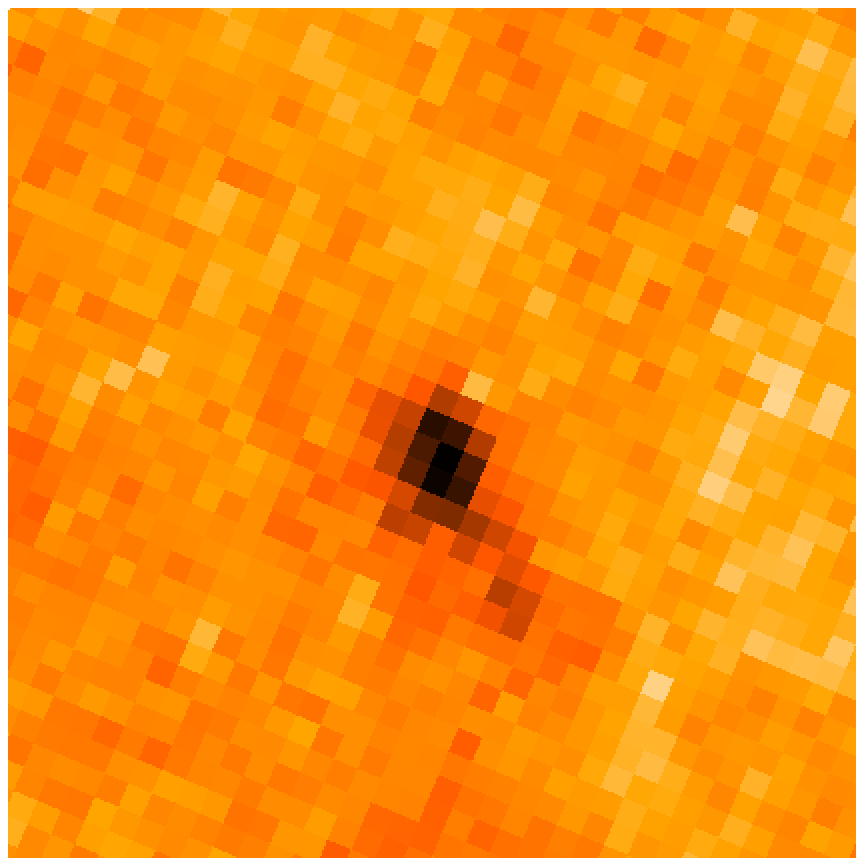}}}
}
\subfigure[HAWK-I {\it J} late-time]{
\includegraphics[width=0.3\textwidth]{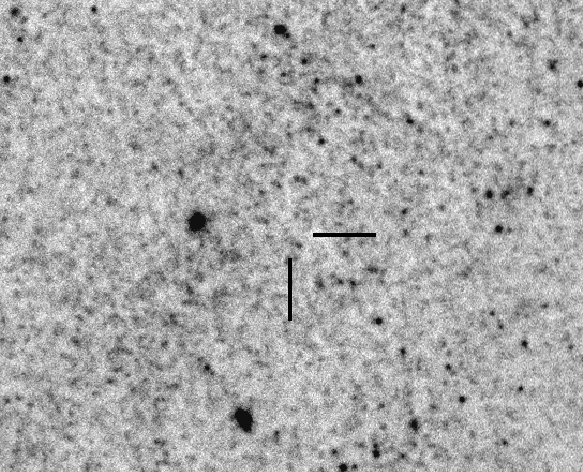}\llap{\raisebox{2.3cm}{\includegraphics[width=2cm,height=2cm]{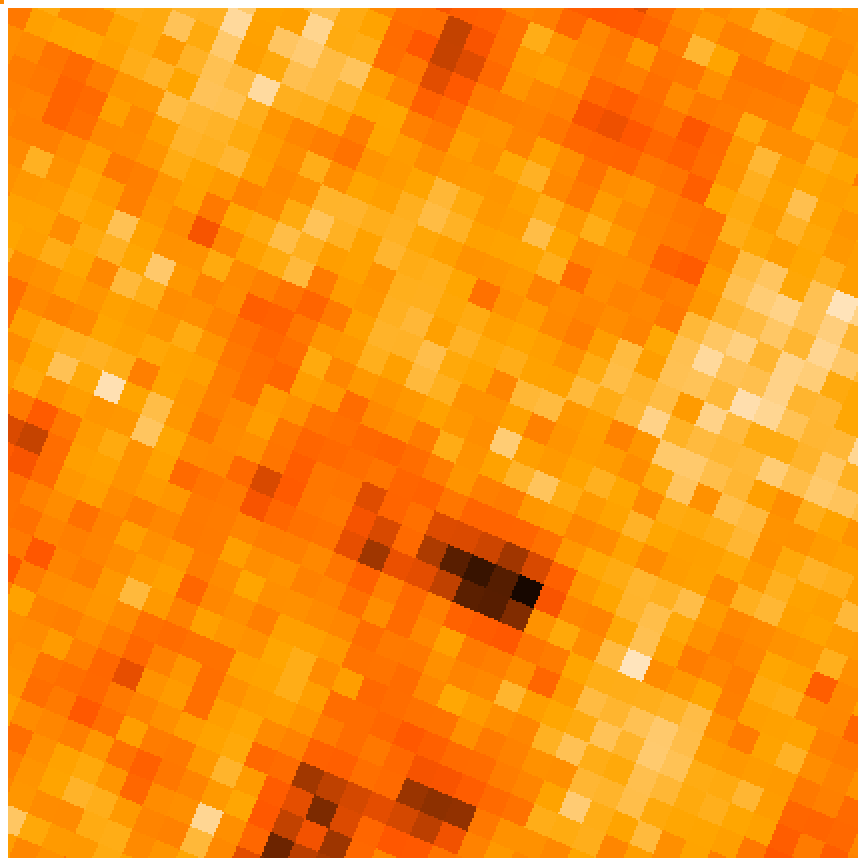}}}
}
\subfigure[Subtracted image ({\it Js})]{
\includegraphics[width=0.3\textwidth]{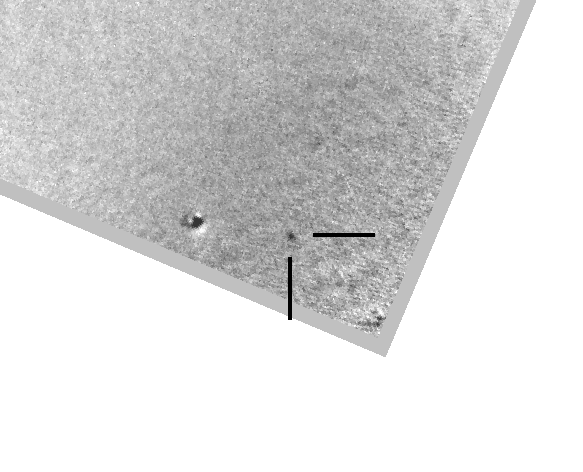}\llap{\raisebox{2.3cm}{\includegraphics[width=2cm,height=2cm]{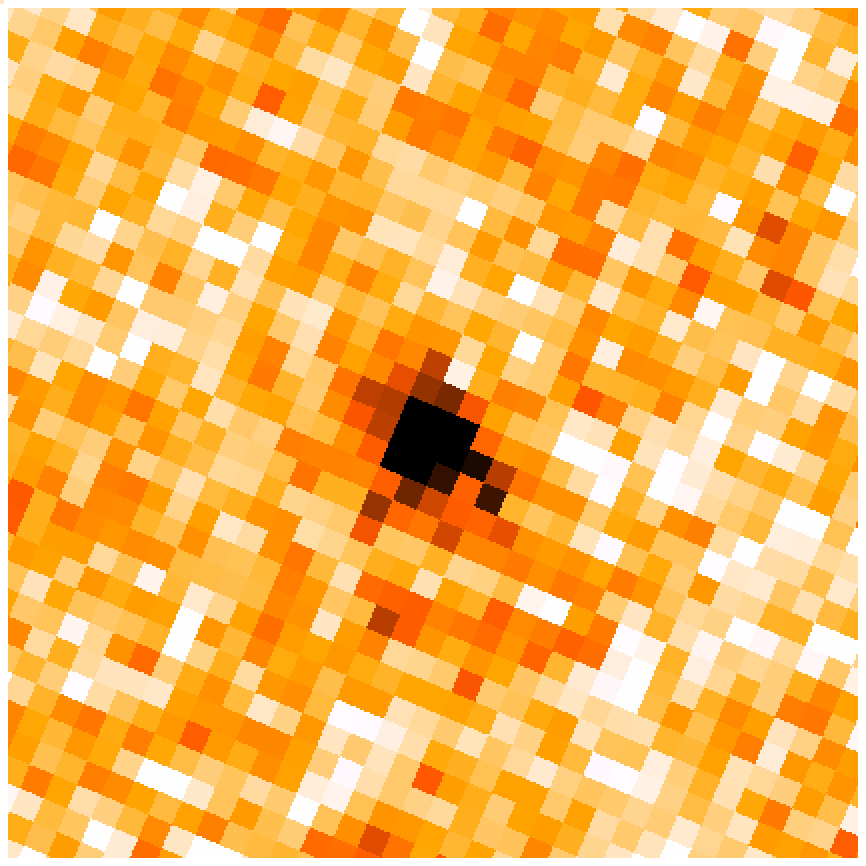}}}
}
 
\subfigure[SOFI {\it Ks} pre-explosion]{
\includegraphics[width=0.3\textwidth]{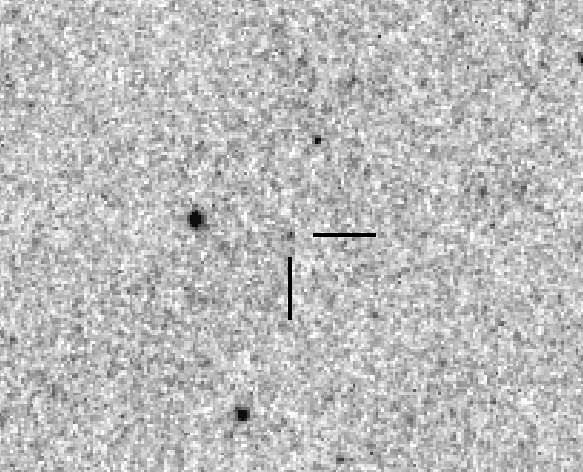}\llap{\raisebox{2.3cm}{\includegraphics[width=2cm,height=2cm]{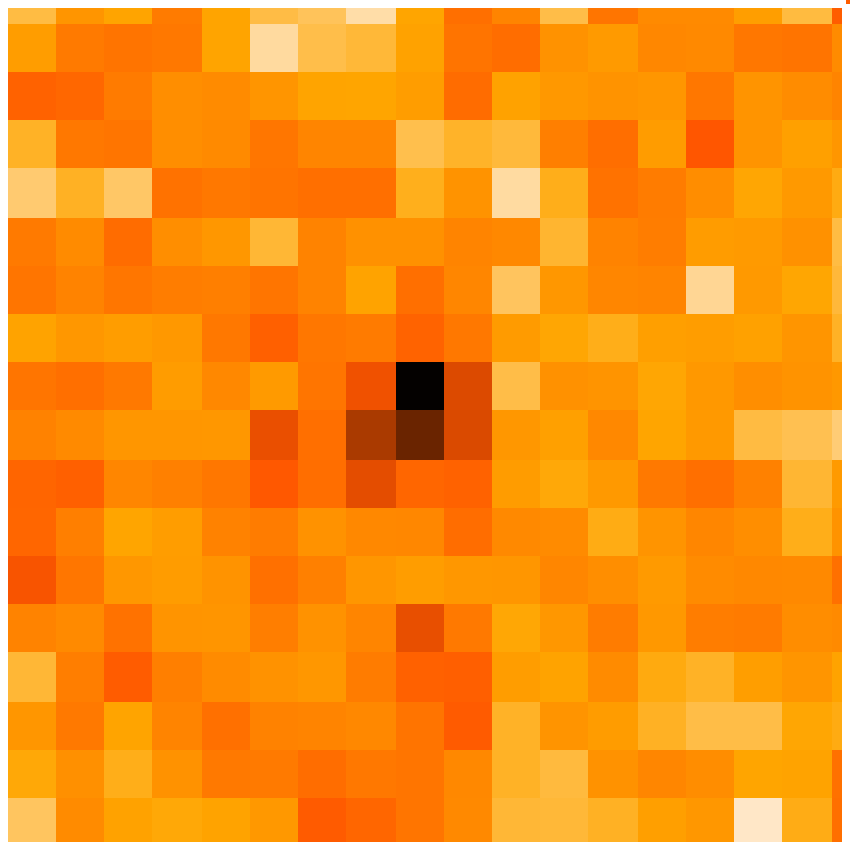}}}
}
\subfigure[HAWK-I {\it Ks} late-time]{
\includegraphics[width=0.3\textwidth]{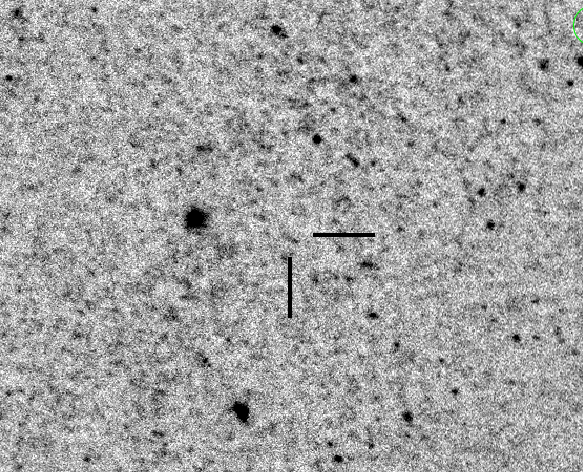}\llap{\raisebox{2.3cm}{\includegraphics[width=2cm,height=2cm]{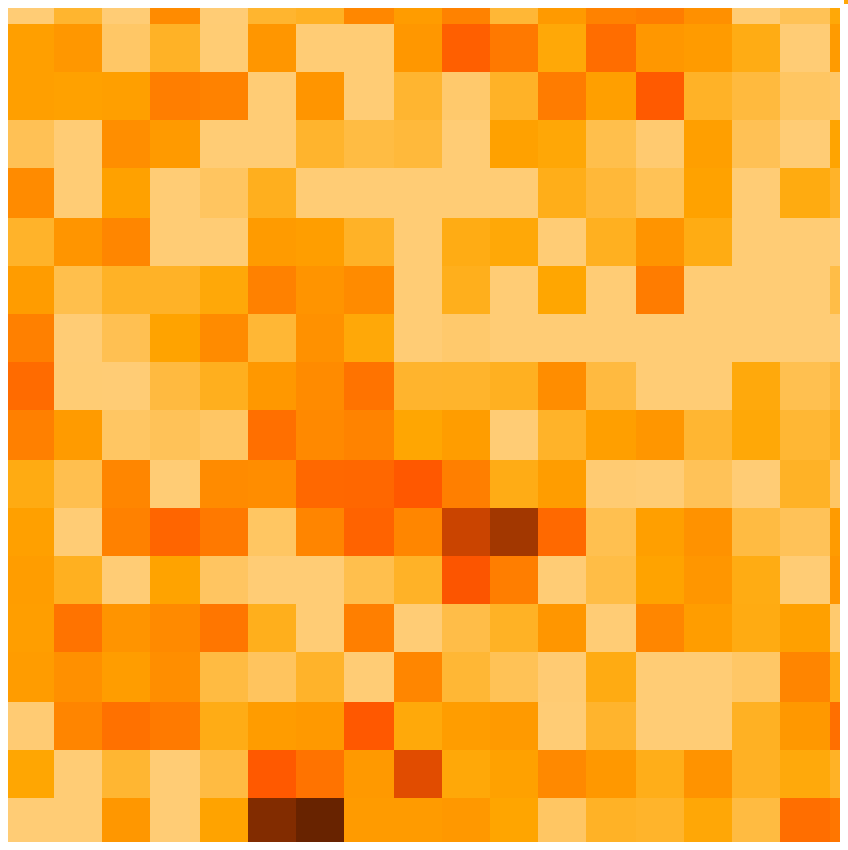}}}
}
\subfigure[Subtracted image ({\it Ks})]{
\includegraphics[width=0.3\textwidth]{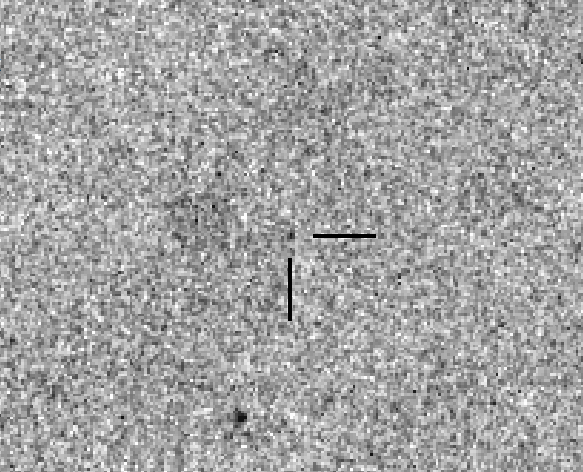}\llap{\raisebox{2.3cm}{\includegraphics[width=2cm,height=2cm]{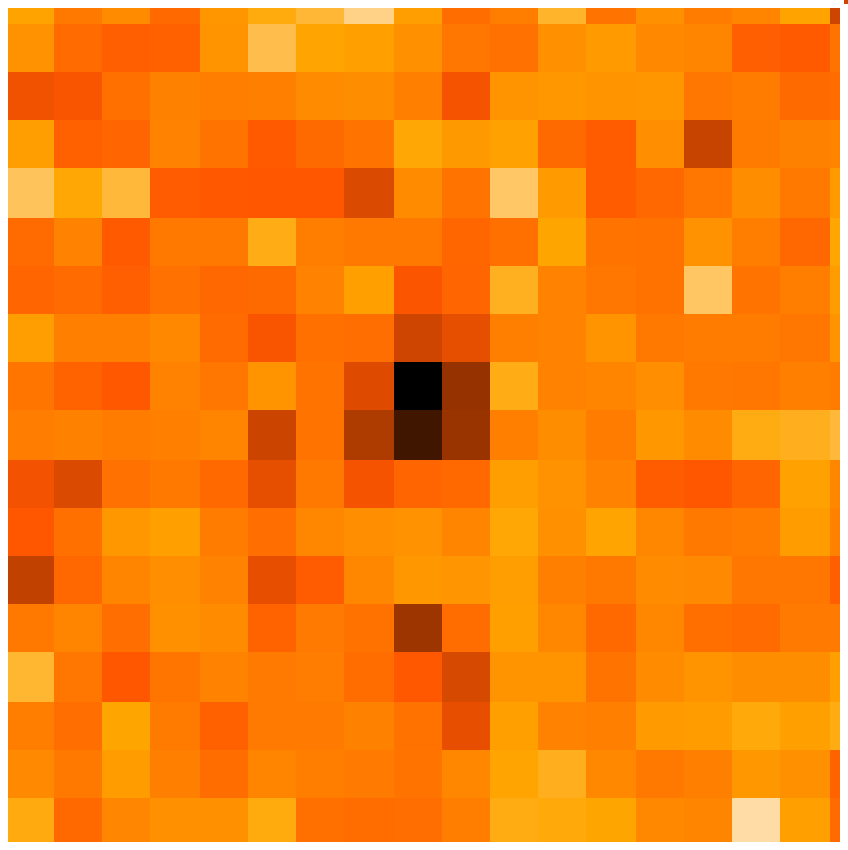}}}
}

\caption[]{The pre-explosion and late-time {\it J} and {\it K}-band images of the site of SN 2012aw, together with the subtracted images. Tick marks indicate the position of SN 2012aw, the scale and orientation of all panels are shown in the upper left. The inset in each panel shows a 5\arcsec$\times$5\arcsec region centred on the SN location. In the inset, the late-time image has been transformed to match the pixel scale and orientation of the pre-explosion image.}
\label{fig:images}
\end{figure*}

To compare the pre-explosion and late-time images, it is necessary to register them to a common pixel grid. The pixel coordinates of point sources common to both frames were measured, and {\sc iraf geomap} was then used to derive a geometric transformation between the matched coordinate lists. This transformation was then applied to the late-time image to map it onto the same pixel grid as the pre-explosion frame. By transforming the late-time rather than the pre-explosion image, we avoid reducing the signal-to-noise of the progenitor candidate. We used the {\sc High Order Transform of PSF And Template Subtraction} ({\sc hotpants}) package to match the Point Spread Functions (PSFs) of the two aligned frames and perform difference imaging between them. The resulting subtracted images are shown in Fig. \ref{fig:images}. In both filters, the source visible in pre-explosion imaging has disappeared at late times, and appears as a source in the difference images, confirming the identification of the progenitor candidate. In the {\it J}-band there is a residual corresponding to a poorly subtracted nearby bright star, and there is some structure to the background in the subtracted image which is caused by proximity to the edge of the image, while in {\it K}-band the subtraction appears good, with only a single source (most likely a variable foreground star) in the difference image apart from the progenitor.

We used eight bright Two-Micron All Sky Survey \citep[2MASS;][]{Skr06} sources within the footprint of the entire HAWK-I mosaic to determine the photometric zeropoint for the HAWK-I {\it Ks}-band image, with an uncertainty in zeropoint of 0.11 mag. These were in turn used to define a local tertiary photometric sequence of intermediate-magnitude sources which were visible in the pre-explosion SOFI {\it Ks} image. The same procedure was used to determine the zeropoint of the {\it J}-band image, with an uncertainty of 0.15~mag.  We performed PSF-fitting photometry on the progenitor in template subtracted images in {\it Ks} and on the un-subtracted image in {\it J} using the {\sc snoopy} package\footnote{{\sc snoopy} in an {\sc iraf} package based on {\sc daophot}, and developed by. E. Cappellaro for SN photometry.}, and calibrated the resulting magnitudes using the local sequence. The magnitude of the progenitor was found to be {\it J}=$20.83\pm0.18$, and {\it Ks}=$19.56\pm0.29$~mag, where the error reflects both the uncertainty in zeropoint, and progenitor photometry as determined by artificial star tests. These magnitudes are consistent with the values found by both \citetalias{Van12} and \citetalias{Fra12}. As an additional test, we also performed aperture photometry on the progenitor in the {\it K}-band image (where the nearby source to the south east is not visible in the pre-explosion frame) and find a magnitude which is consistent to within the uncertainties with the results from PSF-fitting photometry.

In principle, one could improve the photometry for the {\it J}-band pre-explosion data by measuring the magnitude of the progenitor on the template subtracted image, which may reduce the contribution of the nearby source to the south west to the measured flux. However, as we were unable to obtain a satisfactory subtraction for the {\it J}-band image due to the proximity of SN 2012aw to the edge of the detector in the pre-explosion image, measuring its magnitude on the difference image did not improve our results.

\begin{figure}
\includegraphics[width=1\linewidth,angle=0]{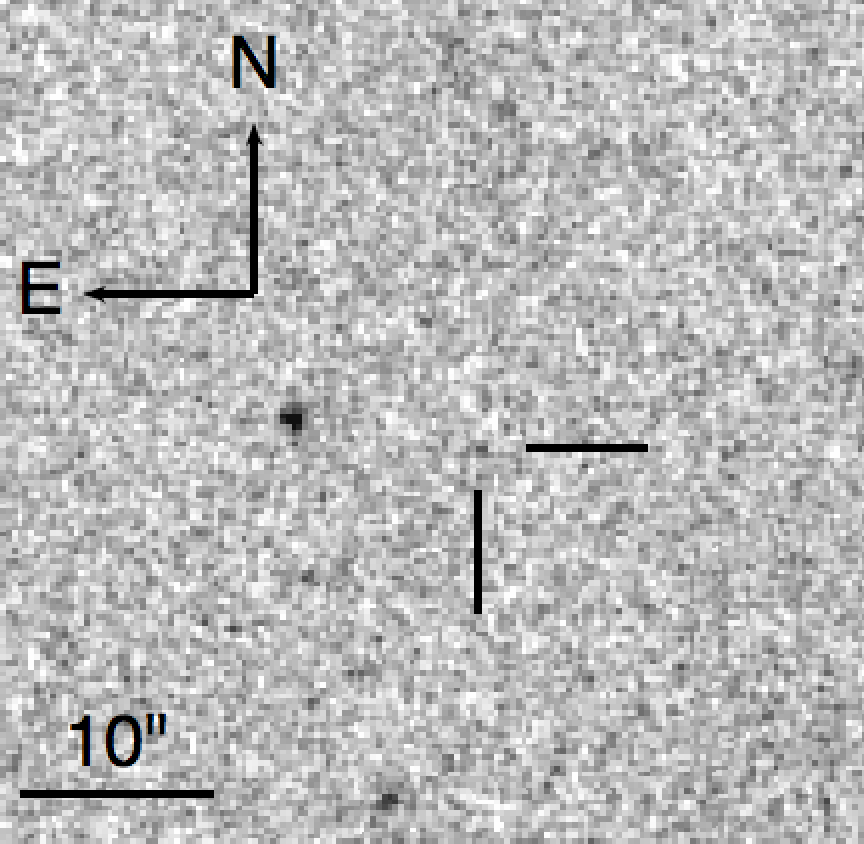}\llap{\raisebox{5.25cm}{\includegraphics[height=3cm]{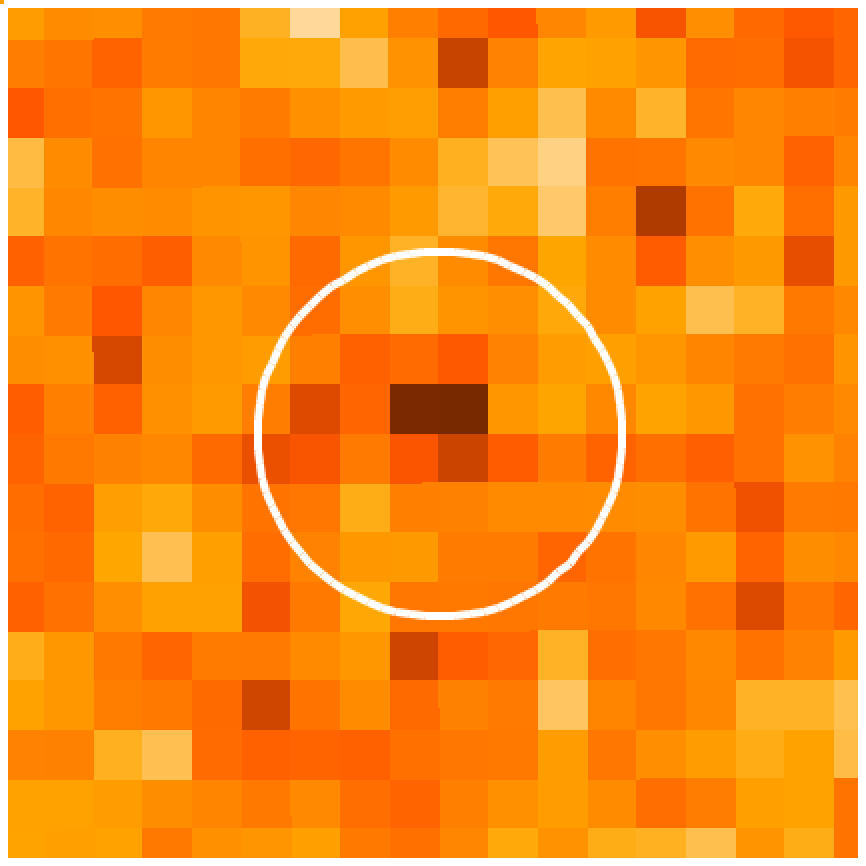}}}
\caption{Pre-explosion SOFI {\it H}-band image showing the detection of the progenitor. Inset in the upper right is a magnified view of the position of SN~2012aw covering 5\arcsec$\times$5\arcsec, with the progenitor circled.}
\label{fig:H}
\end{figure}

As no late-time images were taken in the {\it H}-band, we were unable to perform template subtraction, or to improve the photometric calibration of the pre-explosion data using observations of standard fields. We performed PSF-fitting photometry on the SOFI {\it H} image (shown in Fig. \ref{fig:H}) using {\sc snoopy}, and estimated the photometric error through artificial star tests at the location of the progenitor. The zeropoint of the image was determined through aperture photometry of six sources in the field for which 2MASS photometry were available. While 2MASS has a large photometric uncertainty for faint sources, the uncertainty in zeropoint is insignificant compared to the photometric error for the progenitor. We determine the magnitude of the progenitor to be {\it H}~=~$19.67\pm0.40$.

\section{Discussion}
\label{s4}

Our photometry is consistent with that measured by \citetalias{Fra12} and \citetalias{Van12}, and so the progenitor luminosity calculated by \cite{Koc12} remains valid. We adopt the 1$\sigma$ range for the progenitor luminosity assuming silicate dust from \citeauthor{Koc12}, who constrain the progenitor to have a luminosity of log L/\lsun=4.8--5.0~dex. Using the solar metallicity STARS models \citep{Eld04} as presented in \cite{Sma09}, this luminosity corresponds to a progenitor zero-age main sequence (ZAMS) mass of between 11 and 14~\msun. We hence adopt 12.5$\pm$1.5 \msun\ for the mass of the progenitor of SN 2012aw. We note that other modern stellar evolutionary codes will give similar results, as the final model luminosities for a given mass are comparable to those from the STARS code\citep{Sma09}.

\cite{Van15} presented late-time ultraviolet and optical imaging from {\it HST} taken on 2014 April 23, in which a resolved light echo was observed for SN 2012aw, caused by diffuse interstellar dust scattering the SN flux. The SN itself was still visible in these observations as a point source, although \cite{Van15} found that its magnitude in {\it F814W} was 0.4~mag fainter than that of the progenitor, and suggested that the latter had vanished. The light echo is more pronounced in the bluer bands, and hence will not affect our NIR observations (which were also taken $\sim$1~yr after the {\it HST} data).

The disappearance of the progenitor of SN 2012aw brings the sample of Type IIP SNe for which a red supergiant progenitor candidate has been confirmed through late-time imaging to seven. SN 2012aw is among the highest mass progenitors seen to disappear thus far, and despite some erroneous early claims that it may lie above the observed 16.5~\msun\ limit for Type IIP SN progenitors, it now appears to be securely below this limit. In fact, even if we compare the 1$\sigma$ upper limit on the luminosity of SN 2012aw (log~L/\lsun=5.0~dex), to the model luminosity at the end of core He-burning (which is a conservative {\it lower} limit on the luminosity at the point of core-collapse), then the progenitor must be $<$17~\msun.

Two independent techniques have been applied to estimate the ZAMS mass of the progenitor of SN 2012aw. \cite{Jer14} used nLTE radiative transfer models to derive a progenitor mass from the observed strength of emission lines in the late-time nebular spectra of the SN. Based on this, \citeauthor{Jer14} ruled out a progenitor ZAMS mass $>$20~\msun\, and found that a $\sim$15 \msun\ model consistently reproduced the observations. However, the 12~\msun\ model of \citeauthor{Jer14} also provides a reasonable match to the line strengths, in particular for the sensitive [O~{\sc i}]~$\lambda\lambda$~6300,6364~\AA\ lines. \cite{Dal14} used a radiation-hydrodynamics code to model the early luminosity, temperature and velocity evolution of SN 2012aw, and found a best fitting progenitor model had a 19.6~\msun\ envelope mass. This is clearly inconsistent with the progenitor mass from direct imaging, and moreover continues the tendency for hydrodynamic models to yield higher progenitor masses when compared to direct detections \citep{Utr08}. We note that \cite{Dal14} performed an initial exploration of the potential progenitor parameter space using a semi-analytic code, and that the $\chi^2$ distribution of their models as a function of mass appears to show a local minimum at $\sim$16\msun. However, the full radiation-hydrodynamics code was unable to simultaneously fit the plateau duration and velocities for a model with this mass.

\section{Acknowledgements}

It is a pleasure to thank Seppo Mattila, Rubina Kotak, Nancy Elias-Rosa, Stephen Smartt, Anders Jerkstrand, John Eldridge and Justyn Maund for useful discussions and comments on this work, and Gerry Gilmore for ongoing support.
We thank the anonymous referee for a careful reading of the paper.
Based on observations made with ESO Telescopes at the La Silla Paranal Observatory under programme ID 094.D-0351.
This work was partly supported by the European Union FP7 programme through ERC grant number 320360.

\bsp

\label{lastpage}

\end{document}